\documentclass[showpacs,preprintnumbers,amsmath,amssymb]{revtex4}

\usepackage{graphicx}
\usepackage{dcolumn}
\usepackage{bm}

\begin{document}
\def\bea{\begin{eqnarray}}
\def\eea{\end{eqnarray}}
\def\nn{\nonumber}

\renewcommand\epsilon{\varepsilon}
\def\beq{\begin{equation}}
\def\eeq{\end{equation}}
\def\lla{\left\langle}
\def\rra{\right\rangle}
\def\za{\alpha}
\def\zb{\beta}
\def\lsim{\mathrel{\raise.3ex\hbox{$<$\kern-.75em\lower1ex\h
box{$\sim$}}} }
\def\gsim{\mathrel{\raise.3ex\hbox{$>$\kern-.75em\lower1ex\h
box{$\sim$}}} }
\newcommand{\Rbs}{\mbox{${{\scriptstyle
\not}{\scriptscriptstyle R}}$}}
\newcommand{\cm}{\check{m}}


\title{Minimal seesaw model with tri/bi-maximal mixing and
leptogenesis}

\author{Sanghyeon Chang} \email{schang@phya.yonsei.ac.kr}
\author{Sin Kyu ~Kang} \email{skkang@phya.snu.ac.kr}
\author{Kim ~Siyeon} \email{siyeon@cau.ac.kr}

\affiliation{$^*$Department of Physics, Yonsei University,
        Seoul 120-749, Korea
        \\ $^\dag$School of Physics, Seoul National
University,
        Seoul 151-734, Korea
        \\ $^\ddag$Department of Physics,
        Chung-Ang University, Seoul 156-756, Korea}
\date{\today}

\begin{abstract}
\noindent We examine minimal seesaw mechanism in which the masses
of light neutrinos are described with tri/bi-maximal mixing in the
basis where the charged-lepton Yukawa matrix and heavy Majorana
neutrino mass matrix are diagonal. We search for all possible
Dirac mass textures which contain at least one zero entry in $3
\times 2$ matrix and evaluate the corresponding lepton
asymmetries. We present the baryon asymmetry in terms of a single
low energy unknown, a Majorana CP phase to be clued from
neutrinoless double beta decay.
\end{abstract}


\maketitle \thispagestyle{empty}


\section{Introduction}

\noindent Thanks to the accumulating data from experiments on the
atmospheric and solar neutrinos experiments \cite{atm,SK2002,SNO},
we are now convinced that neutrinos oscillate among three active
neutrinos. Interpreting each experiment in terms of two-flavor
mixing, the mixing angle for the oscillation of atmospheric
neutrinos is understood to be maximal or nearly maximal: $
\sin^2{2\theta_{atm}} \simeq 1,\hspace{4pt} $,
whereas the one for the oscillation of solar neutrinos is not
maximal but large: $ \sin^2\theta_{sol} \simeq 0.3,\hspace{4pt} $
\cite{fits}.
The upper bound for $\theta_{reac}$, $ \sin\theta_{reac} \lesssim
0.2 $, was obtained from the non-observation of the disappearance
of $\overline{\nu_e}$ in the Chooz experiment\cite{chooz} with
$\Delta m^2 \leq 10^{-3} ~\mbox{eV}^2$. The unitary mixing matrix
is defined via
$\nu_{a}=\sum^{3}_{j=1}U_{aj}\nu_{j}~(a=e,\mu,\tau)$, where
$\nu_a$ is a flavor eigenstate and $\nu_j$ is a mass eigenstate.
Including data from SNO\cite{SNO} and KamLand\cite{Eguchi:2002dm},
the range of the magnitude of the MNS mixing matrix is given by
\cite{Guo:2002ei,Fukugita:2003tn,Gonzalez-Garcia:2003qf,
vallefit},
    \bea
    |U| = \left(
        \begin{array}{ccc}
        0.79-0.86 & 0.50-0.61 & 0-0.16 \\
        0.24-0.52 & 0.44-0.69 & 0.63-0.79 \\
        0.26-0.52 & 0.47-0.71 & 0.60-0.77
        \end{array}\right)
    \label{mixnum}
    \eea
at the 90\% confidence level. The existing data also show that the
neutrino mass squared differences which induce the solar and
atmospheric neutrino oscillations are $\Delta m_{sol}^2 \simeq
\left( 7^{+10}_{-2} \right) \times 10^{-5}~\mbox{eV}^2$ and
$\Delta m^2_{atm}\simeq \left( 2.5^{+1.4}_{-0.9} \right) \times
10^{-3}~\mbox{eV}^2$, respectively. It can be readily recognized
that the central values of elements in the mixing matrix in
Eq.(\ref{mixnum}) are pointing an elegant form, which is called
tri/bi-maximal mixing\cite{Harrison:2003aw},
    \bea
    U_{TB} = \left(\begin{array}{ccc}
         \frac{2}{\sqrt{6}}& \frac{1}{\sqrt{3}} & 0 \\
        -\frac{1}{\sqrt{6}}& \frac{1}{\sqrt{3}} &
        -\frac{1}{\sqrt{2}} \\
        -\frac{1}{\sqrt{6}}& \frac{1}{\sqrt{3}} &
        \frac{1}{\sqrt{2}}
    \end{array}\right),
    \label{tribi}
    \eea
which consists of $\sin\theta_{sol}=\frac{1}{\sqrt{3}}$ and
$\sin\theta_{atm}=\frac{1}{\sqrt{2}}$. There are some
literatures\cite{tribi} which proposed textures of the mass matrix
based on the particular mixing type $U_{TB}$.

On the other hand, the baryon density of our universe $\Omega_B
h^2 = 0.0224 \pm 0.0009$ implied by WMAP(Wilkinson Microwave
Anisotropy Probe) data indicates the observed baryon asymmetry in
the Universe\cite{WMAP1, WMAP2},
    \begin{equation}
    \eta_B^{CMB}=
    \frac{n_B-n_{\bar{B}}}{n_\gamma}=
    \left(6.5^{+0.4}_{-0.3}\right) \times 10^{-10},
    \label{baryon}
    \end{equation}
where $n_B, n_{\bar{B}}$ and $n_\gamma$ are number density of
baryon, anti-baryon and photon, respectively. The leptogenesis
\cite{lep} has become a compelling theory to explain the observed
baryon asymmetry in the universe, due to increasing reliance on
the seesaw mechanism from experiments. Theory for lepton asymmetry
requires two heavy right-handed neutrinos or more. For that
reason, a class of models with two heavy right-handed neutrinos
and $3\times 2$ neutrino Dirac mass matrix is called the minimal
neutrino seesaw models(MNSMs) which were intensively studied by
several authors recently \cite{endoh, barger}, especially for
simple Dirac mass textures that make prediction compatible with
solar and atmospheric neutrino data.

The main framework of our work is seesaw mechanism in bottom-up
approach. We launch our analysis by taking $U_{TB}$ for mixing of
light neutrinos and then investigate the structure of $3 \times 2$
Dirac matrix. That is, our concern remains on the combination of
tri/bi-maximal mixing and MNSMs, in order to study the
phenomenological implication of the high energy theory based on
the low energy theory.
One advantage of our framework is that  physical observables can
be explained in minimal terms of physical parameters. In Section
II, we present the light neutrino mass matrix in terms of the
mixing given in Eq.(\ref{tribi}) and mass squared differences
measured in experiment. The mass matrix reconstructed in that way
will constrain Dirac mass matrix. In subsections, depending on the
type of mass hierarchy, possible $3 \times 2$ Dirac matrices will
be examined carefully. In Section III, leptogenesis will be
discussed in details based on the Dirac matrices investigated
before. In section IV, we will present numerical results on
leptogenesis in our scheme and a relationship between leptogenesis
and neutrinoless double beta decay as well as the lower bound of
$M_1$ will be discussed focusing on the contribution from a single
Majorana phase.

\section{Dirac mass matrices in minimal seesaw}

In general, a unitary mixing matrix for 3 generations of neutrinos
is given by
    \begin{eqnarray}
    \tilde{U} &=& R\left(\theta_{23}\right)
          R\left(\theta_{13},\delta\right)
          R\left(\theta_{12}\right)P(\varphi, \varphi')
    \label{fulltrans}
    \end{eqnarray}
where $R$'s are rotations with three angles and a Dirac phase
$\delta$ and the $P=Diag\left(1,e^{i\varphi/2},e^{i\varphi'/2}
\right)$ with Majorana phases $\varphi$ and $\varphi'$ is a
diagonal phase transformation. The mass matrix of light neutrinos
is given by $M_{\nu} = \tilde{U} Diag(m_1,m_2,m_3) \tilde{U}^T$,
where $m_1,m_2,m_3$ are real positive masses of light neutrinos.
Or the Majorana phases can be embedded in the diagonal mass matrix
such that
    \begin{equation}
    M_{\nu} = U Diag(m_1,\cm_2,\cm_3) U^T,
    \label{umu}
    \end{equation}
where $U \equiv \tilde{U}P^{-1}$ and $\cm_2 \equiv
m_2e^{i\varphi}$ and $\cm_3 \equiv m_3e^{i\varphi'}$.

If the $U_{TB}$ in Eq.(\ref{tribi}) is adopted for the $U$ in
Eq.(\ref{umu}), the light neutrino mass is
    \bea
    M_{\nu}
    = m_1\left(\begin{array}{ccc}
      1 & 0 & 0 \\ 0 & 1 & 0 \\ 0 & 0 & 1
\end{array}\right)
      + \frac{\check{m}_2-m_1}{3}\left(\begin{array}{ccc}
      1& 1 & 1 \\ 1 & 1 & 1 \\ 1 & 1 & 1
      \end{array}\right)
      +\frac{\cm_3-m_1}{2}\left( \begin{array}{ccc}
         0 & 0 & 0 \\
         0 & 1 & -1 \\
         0 & -1 & 1 \end{array} \right),\label{mass1}
    \eea
which orients toward a minimal model of neutrino sector by
removing an angle and the Dirac phase. With SNO and KamLand, data
have narrowed down the possible mass spectrum of neutrinos into
two types, Normal Hierarchy ($NH$), $m_1\lesssim m_2 < m_3$, and
Inverse Hierarchy ($IH$), $m_3 < m_1 \lesssim m_2$ for MSW LMA.
Those two types include the possibility of zero mass for a
neutrino, which is necessarily followed by relegating one of the
Majorana phases to the unphysical. In other words, the minimal
model with the physical observables which the present experimental
data guarantee can be obtained by $U_{TB}$ and dictating zero mass
to one generation of neutrinos, where the non-zero physical
parameters in the model consist of 2 masses, 2 angles, one
Majorana phase.

When only two of three neutrinos are massive, by accommodating the
experimental results $\Delta m_{sol}^2=m_2^2-m_1^2$ and $\Delta
m_{atm}^2=|m_3^2-m_2^2|$ to the two types of mass hierarchies, one
can obtain the following expressions for mass eigenvalues,
    \begin{eqnarray}
        \begin{array}{l}
        m_1=0 \\
        m_2=\sqrt{\Delta m_{sol}^2} \\
        m_3=\sqrt{\Delta m_{sol}^2+\Delta m_{atm}^2}
        \\ \end{array} \hspace{10pt}\mbox{for \ $NH$}
        \label{nh} \\ \nonumber \\
            \begin{array}{l}
            m_1=\sqrt{\Delta m_{atm}^2 -
                \Delta m_{sol}^2} \\
            m_2=\sqrt{\Delta m_{atm}^2}\\
            m_3=0
            \label{ih} \\
            \end{array} \hspace{10pt}\mbox{for \ $IH$}.
    \end{eqnarray}
Phase transformation $P=Diag\left(1,e^{i\varphi/2},1 \right)$ now
can replace the phase transformation in Eq.(\ref{fulltrans})
without loss of generality, whether $NH$ or $IH$, so that one can
single $\cm_2$ out in order to investigate the CP violating
contribution of the Majorana phase.

Effective neutrino mass models with one zero mass eigenvalue
involved in three active neutrinos can be generated naturally from
MNSMs. In the basis the mass matrix $M_R$ of right-handed
neutrinos $N_R=(N_1, N_2)$ is diagonal, the model is given
    \begin{equation}
    {\cal L}= -\overline{l_L} M_L l_R
    - \overline{\nu_L} m_D N_R
    +\frac{1}{2} \overline{N^c_R} M_R N_R + h.c.,
    \label{lagrangian}
    \end{equation}
from which the light masses are derived through the seesaw
mechanism, $M_\nu = - m_D M_R^{-1} m_D^T$ in top-down approach. On
the other hand, the matrix $m_D$ is found as the solution to the
seesaw mechanism in bottom-up approach once one launches the
analysis with the light neutrino masses $M_\nu$. Let $M_1$ and
$M_2$ be the masses of right-handed neutrinos and $M_{ij}$ the
elements of the matrix $M_\nu$. The Dirac matrix,
    \begin{equation}
        m_D=\left(
        \begin{array}{cc}
        \sqrt{M_1} a_1 & \sqrt{M_2} b_1 \\
        \sqrt{M_1} a_2 & \sqrt{M_2} b_2 \\
        \sqrt{M_1} a_3 & \sqrt{M_2} b_3
        \end{array} \right),
    \label{dirac}
    \end{equation}
is resulted in with
    \begin{eqnarray}
        && a_1 = \sqrt{M_{11} - b_1^2},~~
        b_1 = \sqrt{M_{11} - a_1^2}, \nonumber \\
        && a_i = \frac{1}{M_{11}} \left[ a_1 M_{1i} -
        \sigma_i b_1 \sqrt{M_{11}M_{ii}-M_{1i}^2} \right],
        \nonumber \\
            && b_i = \frac{1}{M_{11}} \left[ b_1 M_{1i} +
            \sigma_i a_1 \sqrt{M_{11}M_{ii}-M_{1i}^2}\right],
            \label{key} \\
            && M_{11}M_{23} =  \left[M_{12}M_{13} + \sigma_2\sigma_3
            \sqrt{\left(M_{11}M_{22}-M_{12}^2\right)
            \left(M_{11}M_{33}-M_{13}^2\right)} \right].\nonumber
            \label{bargerhe}
    \end{eqnarray}
where the $i$ is 2 or 3, the $\sigma_i$ is a sign $\pm$, and the
sign of $a_1$ is fixed as positive. The solution in eq.
(\ref{bargerhe}) was first derived and formulated in
Ref.\cite{barger}. It is clear that only 5 parameters out of 6;
$a_1, b_1, a_i, b_i,$ can be specified in terms of the elements of
$M_\nu$. There are various ways to decrease the number of
parameters in Dirac matrix, posing one or more zeros or posing
equalities between elements for the matrix texture. It is known
that texture zeros or equalities among matrix entries can be
generated by imposing additional symmetries to the theory.

In this paper, we focus on posing one or more zeros in Dirac
matrix, and show that only one-zero textures are allowed for $NH$
and only one-zero and two-zero textures are allowed for $IH$,
accompanied with the low energy mixing $U_{TB}$. On the other
hand, from Eq.(\ref{key}), one can recognize that, if there exists
a kind of symmetry between entries in $M_\nu$, the Dirac matrix
also has a symmetry in certain entries inherited from the symmetry
of the $M_\nu$. So, there are a number of patterns with equalities
among the entries in Dirac matrices obtained with one or two
zeros, as a consequence of maximal mixing.

\subsection{Normal Hierarchy}

With $m_1=0$, the neutrino mass $M_\nu$ in Eq.(\ref{mass1})
reduces to
    \begin{equation}
        M_{\nu} =
        \frac{\cm_2}{3}\left(\begin{array}{ccc}
        1 & 1 & 1 \\  & 1 + d & 1 - d \\  &  & 1 + d
        \end{array} \right)
        \label{nhmass},
    \end{equation}
where $d =  3 m_3 / 2 \cm_2$, which, using Eq.(\ref{key}), gives
rise to Dirac matrix with the following entries:
    \begin{eqnarray}\begin{array}{lll}
        a_1= \sqrt{\cm_2 /3 - b_1^2},& &
        b_1= \sqrt{\cm_2 /3 - a_1^2}, \\
        a_i= a_1 - \sigma_i b_1 \sqrt{d},& &
        b_i= b_1 + \sigma_i a_1 \sqrt{d}, \hspace{10pt} i=2,3
        \end{array}
        \label{nhdirac}
    \end{eqnarray}
where $\sigma_2\sigma_3 = -1$. Depending on the position of
texture zero, the types of Dirac matrices can be classified as
follows;
    \begin{itemize}
        \item{$NH$ 1-a : $b_1=0,\hspace{4pt} a_1=\sqrt{\cm_2/3},\hspace{4pt}
         a_1=a_2=a_3,\hspace{4pt} b_2=-b_3=\sigma_2 \sqrt{m_3/2}$}
        \item{$NH$ 1-b : $a_1=0,\hspace{4pt} a_1 \leftrightarrow b_1,
        \hspace{4pt} a_i \leftrightarrow b_i$ in $NH$ 1-a}
        \item{$NH$ 2-a : $b_2=0,\hspace{4pt} a_2=\sqrt{\cm_2/3 +m_3/2},
        \hspace{4pt} a_1=a_2/(1+d),\hspace{4pt} a_3=a_2(1-d)/(1+d),
        \hspace{4pt} b_1/a_1=-\sigma_2 \sqrt{d},\hspace{4pt} b_3=2b_1$}
        \item{$NH$ 2-b : $a_2=0,\hspace{4pt} a_1 \leftrightarrow b_1,
        \hspace{4pt} a_i \leftrightarrow b_i$ in $NH$ 2-a}
    \end{itemize}
The matrix in Eq.(\ref{nhmass}) features the equalities between
$M_{22}$ and $M_{33}$ and between $M_{12}$ and $M_{13}$ as a
consequence of the maximal mixing of atmospheric neutrinos. In the
case with $b_2=0$, it can be recognized that the ratios of $a_1$
to $a_2$ and $a_3$ to $a_2$ inherit those of $M_{12}$ to $M_{22}$
and $M_{23}$ to $M_{22}$, respectively. $b_3=0$ or $a_3=0$ cases
will not be presented as an independent case,
 since it can be made by
exchanging $b_3$ with $b_2$ and $a_3$ with $a_2$ from $NH$ 2-a and
2-b.

\subsection{Inverse Hierarchy}
With $m_3=0$,
    \begin{equation}
        M_{\nu}
        = \frac{m_1}{3}\left(\begin{array}{ccc}
        x+2 & x-1 & x-1 \\  & x + 1/2 & x +1/2  \\  &  &
        x+1/2 \end{array} \right),
    \label{ihmass}
    \end{equation}
where $x\equiv  \check{m}_2/ m_1$, which, using Eq.(\ref{key}),
gives rise to Dirac matrix with the following entries:
    \begin{eqnarray}\begin{array}{lll}
        a_1= \sqrt{m_1 (x+2)/3 - b_1^2},& &
        b_1= \sqrt{m_1 (x+2)/3 - a_1^2}, \\
        a_i= \left((x-1)a_1 - 3\sigma b_1
             \sqrt{x/2}\right)/(x+2), & &
        b_i= \left((x-1)b_1 + 3\sigma a_1
             \sqrt{x/2}\right)/(x+2), \hspace{10pt} i=2,3
        \end{array}
        \label{ihdirac}
    \end{eqnarray}
where $\sigma \equiv \sigma_i$ and $\sigma_2\sigma_3=1$. The
equality $M_{22}=M_{23}=M_{33}$ in Eq.(\ref{ihmass}), which is
again a consequence of the maximal mixing of atmospheric
neutrinos, constrains the elements of the Dirac matrix such that
$a_2=a_3,\hspace{4pt} b_2=b_3$. Hence, texture with a single zero
included appears only if $a_1=0$ or $b_1=0$, while texture with
two zeros appears if $a_2=a_3=0$ or $b_2=b_3=0$ .
\begin{itemize}
    \item{$IH$ 1-a : $b_1=0$,\hspace{4pt} $a_1=\sqrt{\cm_2/3+2m_1/3},
    \hspace{4pt} a_i=a_1(x-1)/(x+2),\hspace{4pt}
    b_i/a_i = 3 \sigma \sqrt{x/2} /(x-1)$}
    \item{$IH$ 1-b : $a_1=0,\hspace{4pt} a_1 \leftrightarrow b_1,
    \hspace{4pt} a_i \leftrightarrow b_i$ in $IH$ 1-a}
    \item{$IH$ 2-a : $b_2=b_3=0,\hspace{4pt} a_2=a_3=\sqrt{\cm_2/3
+ m_1/6},
    \hspace{4pt} a_1=2a_2(x-1)/(2x+1),
    \hspace{4pt} b_1/a_1 = -3 \sigma \sqrt{x/2} /(x-1)$}
    \item{$IH$ 2-b : $a_2=a_3=0,\hspace{4pt} a_1 \leftrightarrow b_1,
    \hspace{4pt} a_i \leftrightarrow b_i$ in $IH$ 2-a}
\end{itemize}

Listed are all the cases with one or more texture zeros in Dirac
matrix derivable from the light neutrino mass with $U_{TB}$,
whether $NH$ or $IH$. In the following, the eligibility of each
case to generate the CP asymmetry for leptogenesis will be
examined.

\section{Leptogenesis}

The baryon asymmetry Eq.(\ref{baryon}) can be rephrased
    \begin{equation}
    Y_B = \frac{n_B-n_{\bar{B}}}{s} \simeq
    \left(8.8 - 9.8 \right) \times 10^{-11}. \label{cosm}
    \end{equation}
The $n_{\gamma}$ is the photon number density and the $s$ is
entropy density so that the number density with respect to a
co-moving volume element is taken into account. The baryon
asymmetry produced through sphaleron process is related to the
lepton asymmetry \cite{Harvey:1990qw,Kolb:vq} by
\bea %
Y_B = a Y_{B-L} = \frac{a}{a-1} Y_L ,%
\eea %
where $ a \equiv (8 N_F + 4 N_H) / (22 N_F + 13 N_H)$, for
example, $a=28/79$ for the Standard Model(SM) with three
generations of fermions and a single Higgs doublet, $N_F = 3, N_H
= 1$. The purpose of this work is to estimate whether the Yukawa
interaction which produces the light neutrinos with the mixing
Eq.(\ref{tribi}) through  the seesaw mechanism can also generate a
sufficient lepton asymmetry for the observed baryon asymmetry.
\noindent The generation of a lepton asymmetry requires the
CP-asymmetry and out-of-equilibrium condition. The $Y_L$ is
explicitly parameterized by two factors, $\epsilon$, the size of
CP asymmetry, and $\kappa$, the dilution factor from washout
process.
\bea %
Y_L = \frac{(n_L - n_{\overline{L}})}{s} = \kappa
\frac{\epsilon_i}{ g^*} \label{aalepto}
\eea %
where $g^*\simeq 110$ is the number of relativistic degree of
freedom. The $\epsilon_i$ is the magnitude of CP asymmetry in
decays of heavy Majorana neutrinos \cite{Kolb:qa,Luty:un},
\bea %
\epsilon_i
        &=&\frac{\Gamma (N_i \to \ell H)
            - \Gamma (N_i \to \bar{\ell} H^*)}
        {\Gamma (N_i \to \ell H)
            + \Gamma (N_i \to \bar{\ell} H^*)},
\label{aacp}
\eea %
where $i$ denotes a generation. When one of two generations of
right neutrinos has a mass far below that for the other
generation, i.e., $M_1 < M_2$, the $\epsilon_i$ in Eq.(\ref{aacp})
is obtained from the decay of
$M_1$\cite{Flanz:1994yx,Covi:1996wh,Buchmuller:1997yu},
\bea %
    \epsilon_1 &=& \frac{1}{8\pi v^2}
            \frac{{\rm Im}\left [({m_D}^\dagger m_D)_{12}^2
\right ]
            } {({m_D}^\dagger m_D)_{11}}
            f\left(\frac{M_2}{M_1}\right)\;, \label{cp1}
\eea %
where $v=174$ GeV and $f\left(M_2/M_1\right)$ represents loop
contribution to the decay width from vertex and self energy and is
given by
\begin{equation}
f(x) = x\left[1-(1+x^2)\ln \frac{1+x^2}{x^2} +
\frac{1}{1-x^2}\right]
\end{equation}
for the Standard Model. For large value of $x$, the leading order
of $f(x)$ is $(-3/2)x^{-1}$.

It is convenient to consider separately the factor that depends on
Dirac matrix in $\epsilon_1$ in Eq.(\ref{cp1}) at this stage.
    \begin{equation}
        \frac{{\rm Im}\left [({m_D}^\dagger m_D)_{12}^2\right ]
        } {({m_D}^\dagger m_D)_{11}} =
        M_2\frac{Im \left[( a_1^* b_1 + a_2^* b_2+a_3^* b_3)^2\right]}
        {|a_1|^2 + |a_2|^2 +|a_3|^2 }\equiv M_2 \Delta_1,
        \label{delta1}
    \end{equation}
where $a$'s and $b$'s are defined in Eq.(11). From a number of
types of matrices with a texture zero derived in
Eq.(\ref{nhdirac}) and Eq.(\ref{ihdirac}), only 6 different
non-zero values of $\Delta_1$'s can be evaluated. Those particular
Dirac matrices to contribute the imaginary parts are the matrix
with $b_2=0$ and that with $a_2=0$ for $NH$, and the matrix with
$b_1=0$, that with $a_1=0$, that with $b_2=b_3=0$, and that with
$a_2=a_3=0$ for $IH$. For $NH$, if $a_1=0$, or $b_1=0$, the
$(m_D^\dagger m_D)_{12}$ vanishes from the trivial relation
between entries. Applying Eq.(\ref{nhdirac}) and
Eq.(\ref{ihdirac}) for Eq.(\ref{delta1}), one can find that each
type of Dirac matrix gives rise to $\Delta_1$ as follows;
    \begin{eqnarray}
        \Delta_1(NH \hspace{4pt} 2-a) &=&
        \frac{6m_2m_3(m_3^2-m_2^2)\sin\varphi}
        {(2m_2^2+3m_3^2)\sqrt{4m_2^2+9m_3^2+12m_2m_3\cos\varphi}},
        \label{Delta1} \\
        \Delta_1(NH \hspace{4pt} 2-b) &=&
        \frac{-6(m_3^2-m_2^2)\sin\varphi}
        {5\sqrt{4m_2^2+9m_3^2+12m_2m_3\cos\varphi}},
    \end{eqnarray}
where $m_2$ and $m_3$ are given in terms of $\Delta m_{sol}^2$ and
$\Delta m_{atm}^2$ in Eq.(\ref{nh}),
    \begin{eqnarray}
        \Delta_1(IH \hspace{4pt} 1-a) &=&
        \frac{-2m_1m_2(m_2^2-m_1^2)\sin\varphi}
        {(2m_1^2+m_2^2)\sqrt{4m_1^2+m_2^2+4m_1m_2\cos\varphi}}, \\
        \Delta_1(IH \hspace{4pt} 1-b) &=&
        \frac{2(m_2^2-m_1^2)\sin\varphi}
        {3\sqrt{4m_1^2+m_2^2+4m_1m_2\cos\varphi}}, \\
        \Delta_1(IH \hspace{4pt} 2-a) &=&
        \frac{-2m_1m_2(m_2^2-m_1^2)\sin\varphi}
        {(m_1^2+2m_2^2)\sqrt{m_1^2+4m_2^2+4m_1m_2\cos\varphi}}, \\
        \Delta_1(IH \hspace{4pt} 2-b) &=&
        \frac{2(m_2^2-m_1^2)\sin\varphi}
        {3\sqrt{m_1^2+4m_2^2+4m_1m_2\cos\varphi}},
        \label{delta6}
    \end{eqnarray}
where $m_1$ and $m_2$ are given in terms of $\Delta m_{sol}^2$ and
$\Delta m_{atm}^2$ in Eq.(\ref{ih}). Thus, for $M_2 \gg M_1$ case,
the CP asymmetry in Eq.(\ref{cp1}) reduces to $\epsilon_1 \approx
3/(16\pi v^2) M_1 \Delta_1$, which is now parameterized by the
lightest mass of heavy neutrino $M_1$ and Majorana phase
$\varphi$. The sign of $\epsilon_1$ depends on the position of a
texture zero in a row of Dirac matrix.


The $\kappa$ in Eq.(\ref{aalepto}) is determined by solving the
full Boltzmann equations. The $\kappa$ can be simply parameterized
in terms of $K$ defined as the ratio of $\Gamma_1$ the tree-level
decay width of $N_1$ to $H$ the Hubble parameter at temperature
$M_1$, where $K\equiv \Gamma_1 / H<1$ describes processes out of
thermal equilibrium and $\kappa<1$ describes washout
effect\cite{Kolb:vq}\cite{boltz},
\bea %
    \kappa \simeq \frac{0.3}{K \left(\ln K \right)^{0.6}}
    &\rm{for}& 10 \lesssim K \lesssim 10^6, \\
        \kappa \sim \frac{1}{2 \sqrt{K^2+9}}
        &\rm{for}& 0 \lesssim K \lesssim 10.
\eea %
The decay width of $N_1$ by the Yukawa interaction at tree level
and Hubble parameter in terms of temperature $T$ and the Planck
scale $M_{pl}$ are $\Gamma_1 =  \left(m_D^\dagger m_D
        \right)_{11} M_1 / (8 \pi v^2) $ and $H = 1.66 g^{1/2}_* T^2 /
        M_{pl}$, respectively.
At temperature $T = M_1$, the ratio $K$ is
    \begin{eqnarray}
    K = \frac{M_{pl}}{1.66 \sqrt{g^*}(8 \pi v^2)}
    \frac{(m_D^\dagger m_D)_{11}}{M_1},
    \end{eqnarray}
which reduces to, using the Dirac matrices in Eq.(\ref{dirac}),
    \begin{eqnarray}
    K \approx \frac{1}{10^{-3} \rm{eV}}
        \left( |a_1|^2 + |a_2|^2 + |a_3|^2 \right),
    \label{kay}
    \end{eqnarray}
where all fixed numbers are included in a factor of order. As done
for the $\Delta_1$'s, one can apply Eq.(\ref{nhdirac}) and
Eq.(\ref{ihdirac}) for Eq.(\ref{kay}) to find dilution factor
$\kappa$ when the decay width is determined by Yukawa couplings in
each type of Dirac matrix. For the six types of Dirac matrices
that are eligible for the CP asymmetry as in
Eqs.(\ref{Delta1})-(\ref{delta6}), the ratio $K$ for each case is
    \begin{eqnarray}
    K (NH \hspace{4pt} 2-a,\hspace{5pt} -b) &\approx&
    \frac{\left(\begin{array}{lll}
    2m_2^2+3m_3^2, & & 5m_2m_3
    \end{array}\right)}
    {(10^{-3} \rm{eV}) \sqrt{4m_2^2+9m_3^2+12m_2m_3\cos{\varphi}}}\\
        K (IH \hspace{4pt} 1-a,\hspace{5pt} -b) &\approx&
        \frac{\left(\begin{array}{lll}
        2m_1^2+m_2^2, & & 3m_1m_2
        \end{array}\right)}
        {(10^{-3} \rm{eV}) \sqrt{4m_1^2+m_2^2+4m_1m_2\cos{\varphi}}}\\
            K (IH \hspace{4pt} 2-a,\hspace{5pt} -b) &\approx&
            \frac{\left(\begin{array}{lll}
            m_1^2+2m_2^2, & & 3m_1m_2
            \end{array}\right)}
            {(10^{-3} \rm{eV}) \sqrt{m_1^2+4m_2^2+4m_1m_2\cos{\varphi}}},
    \end{eqnarray}
which shows that the dilution factor also depends on the phase
$\varphi$, but it does not significantly affect the order of
magnitude. Out of all the types of Dirac matrices examined, there
is no such a case that Yukawa couplings originate decays of
neutrinos $N_1$ which satisfy the out-of-equilibrium condition
$K<1$ at $T=M_1$. The washout effect of asymmetry is most
suppressed with the Dirac matrix of type $NH\hspace{4pt}2-b$,
where, depending on $\varphi$, the dilution factor ranges from
0.010 to 0.013, the amount of asymmetry survived from washout is
at most about 1$\%$. When $T<M_1$, the Boltzmann equations still
depict the finite value of $\kappa$ as $M_1/T$ increases for the
universe evolution
\cite{Kolb:qa}\cite{Luty:un}\cite{Buchmuller:2000as}.

\section{Discussion}
Based on the formulation of the leptogenesis derived in the
previous section, we numerically analyze baryon asymmetry for each
case classified as $NH$ or $IH$. For the numerical calculation, we
take $\Delta m_{sol}^2=7.0\times 10^{-5}~\mbox{eV}^2$ and $\Delta
m_{atm}^2=2.5\times 10^{-3}~\mbox{eV}^2$ as inputs.

Consider a model with neutrino masses in normal hierarchy. In
Fig.1, we plot the baryon asymmetry $Y_B$ as a function of the
Majorana phase $\varphi$  for $NH 2-b$. The different curves
correspond to $M_1=2.0\times 10^{11}$ to $2.0\times10^{13}$ GeV
for fixed $M_2/M_1=5$. We note that we can choose any reasonable
$M_2/M_1$ value which can protect L-violating processes with $N_1$
from the wash-out when $T<M_2$. As expected from Eq.(\ref{cp1}),
the value of $Y_B$ for a fixed $\varphi$ increases with $M_1$. The
horizontal line in Fig.1 presents the current cosmological
observation of $Y_B$ given in Eq.(\ref{cosm}). From the analysis,
we see that the current observation on $Y_B$ constrains the lower
bound of $M_1$, which turns out to be $M_1\gtrsim 2.0\times
10^{11}$ GeV.  It is clear that the CP asymmetry in high energy is
almost proportional to the imaginary part of Majorana CP
contribution in low energy from Eqs.(\ref{delta1})-(\ref{delta6}).
Thus, the plots show that the lower bound of $M_1$ to generate the
observed baryon asymmetry should be raised if the imaginary
contribution of low energy phase is decreased as the $\varphi$
approaches $0$ or $\pi$. In all aspects of the prediction of
$Y_B$, $NH 2-a$ and $NH 2-b$ are quite similar to each other
except an overall factor. The $Y_B$ for $NH 2-b$ is enhanced from
both the enhancement of CP asymmetry,
$\Delta_1(b)/\Delta_1(a)\simeq 3.6$, and the suppression of
wash-out effect, $\kappa(b)/\kappa(a)\simeq 4.5$. The lower bound
of $M_1$ with $\varphi=\pi/2$ is pulled down to $2.0 \times
10^{11}$ GeV for $NH 2-b$, whereas that for $NH 2-a$ is $3.2
\times 10^{12}$GeV. Suppressing a certain Yukawa coupling by
putting a texture zero can vary the amount of the asymmetry by
order of magnitude.

In Fig.2, we plot $Y_B$ as a function of the Majorana phase
$\varphi$ for $IH1-a$. The different curves correspond to
$M_1=5.5\times 10^{13}$ to $5.5 \times 10^{15}$ GeV for fixed
$M_2/M_1=5$. As in $NH$, we obtain a lower bound on $M_1\gtrsim
5.5\times 10^{12}$ GeV for $IH$. The prediction of $Y_B$ for $IH$
with the same value of $M_1$ is smaller than that for $NH$ because
$\Delta_1$ for $IH$ is proportional to $m_2^2-m_1^2$ which
corresponds to the solar mass squared difference, while $\Delta_1$
for $NH$ is proportional to $m_3^2-m_2^2$ which corresponds to the
atmospheric mass squared difference. We expect from Eqs.(25-28)
that the predictions of $Y_B$'s for other cases of $IH$ are almost
the same as that for $IH$ 1-a because $m_1m_2/(2m_1^2+m_2^2)\sim
1/3$.

Although the Majorana phase is not detectable through neutrino
oscillations, it may affect the amplitude of neutrinoless double
beta decay. Thus, one can anticipate that there may exist a
correlation between leptogenesis and neutrinoless double beta
decay in our scenario where the heavy Majorana neutrino mass
matrix and the charged-lepton Yukawa matrix are both diagonal. The
neutrinoless double beta decay amplitude is proportional to the
effective Majorana mass $|\langle m_{ee} \rangle|$ which can be
written in the form:
\begin{eqnarray}
|\langle m_{ee} \rangle| &=& |\sum_{i=1}^3 U_{ei}^2 m_i
e^{i\varphi_i}|\nonumber \\
&=&\left\{\begin{array}{c} \frac{m_2}{3},~~~~(NH) \\
  \frac{1}{3}(4m_1^2+m_2^2+4m_1m_2\cos\varphi)^{1/2}
,~~~~(IH)\\
\end{array}  \right.\label{bb}
\end{eqnarray}
where $\varphi_i$ are Majorana CP-violating phases. The $|\langle
m_{ee} \rangle|$ depends on the CP phase $\varphi$ only with
inverted hierarchy, so that one can draw a simple correlation
between leptogenesis and neutrinoless double beta decay only for
the particular case. In Fig.3, we present a correlation between
$Y_B$ and $|\langle m_{ee} \rangle|$ for $IH$ 1-a. The inputs are
taken to be the same as in Fig.2. As the value of $|\langle
m_{ee}\rangle|$ approaches to that with $\varphi=\pi/2$, the
asymmetry is enhanced and the bound of $M_1$ becomes lower. The
lower bound of $M_1$ as a function of Majorana phase or that of
effective Majorana mass is obtained from the current cosmological
observation of $Y_B$. In Fig.4, we present a correlation between
the lower bound of $M_1$ and $|\langle m_{ee} \rangle|$.

We examined the minimal seesaw mechanism of $3 \times 2$ Dirac
matrix by starting our analysis with the masses of light neutrinos
with tri/bi-maximal mixing in the basis where the charged-lepton
Yukawa matrix and heavy Majorana neutrino mass matrix are
diagonal. We found all possible Dirac mass textures which contain
one zero entry or two in the matrix and evaluated the
corresponding lepton asymmetries. The baryon asymmetry can be
presented in terms of low energy observables, where only one
Majorana CP phase among them remains yet unknown. The numerical
work exhibits the dependence of both the size of baryon asymmetry
and the lower bound of $M_1$ upon the low energy CP phase to be
clued from neutrinoless double beta decay.


{\it Note added :} After completing this wrok, we have been
noticed that similar analysis for the hierarchical case in
supersymmetric seesaw model appeared in Ref.\cite{ibarra}.

\begin{acknowledgments}
S.K.K is supported in part  by BK21 program of the Ministry of Education
in Korea and in part by KOSEF Grant No. R01-2003-000-10229-0. S.C. is supported in part by Grant No.
R02-2003-000-10050-0 from BRP of the KOSEF and by Brainpool
program of the KOFST. K.S. was supported by the Basic Science
Research Institute Special Program of Chung-Ang University in
2004.
\end{acknowledgments}

\begin{figure}
        \begin{center}
        \includegraphics[width=16cm]{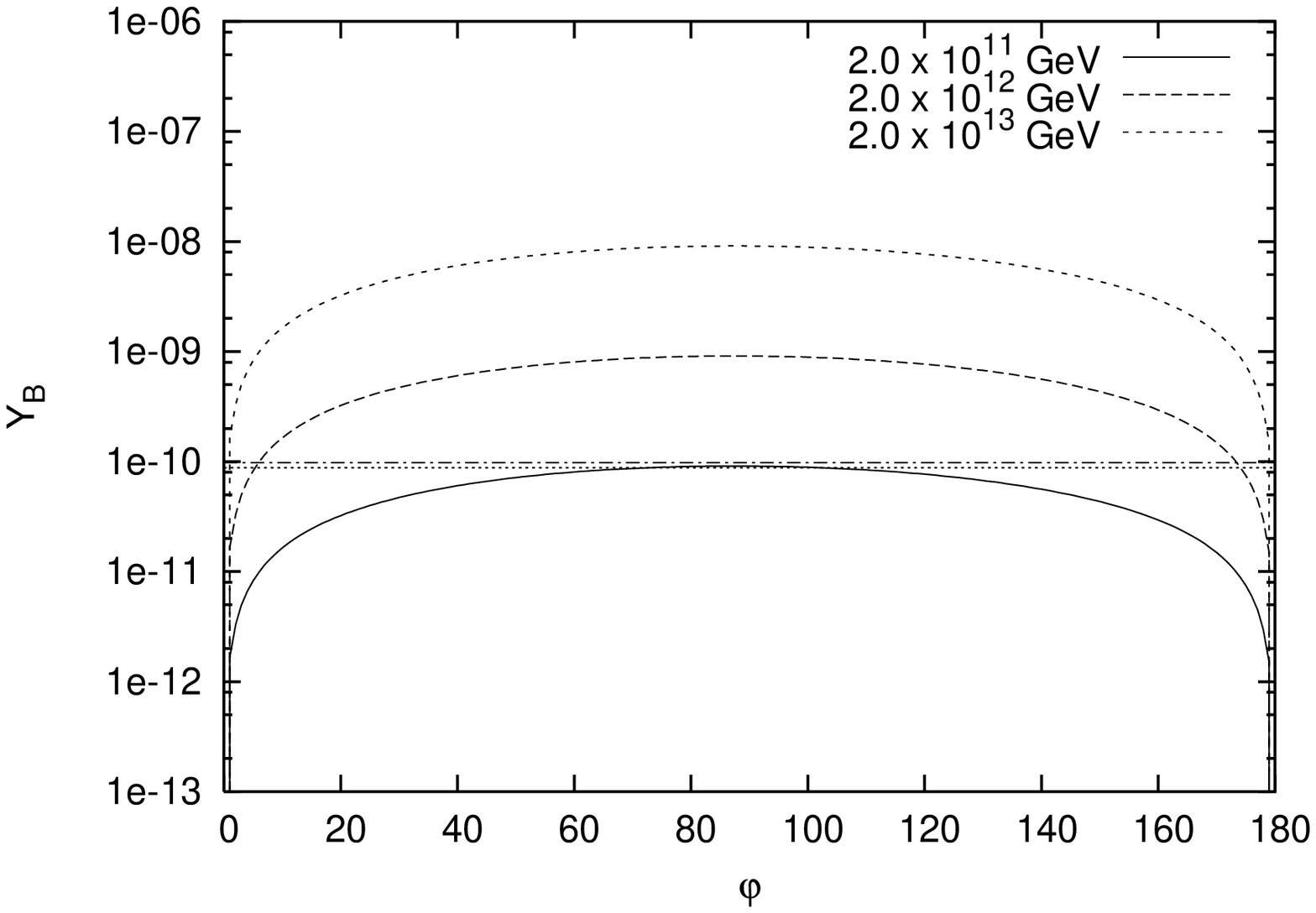}
        \end{center}
    \caption{$Y_B$ as a function of Majorana CP phase for case $NH$
    2-b, with various values of $M_1$ where $M_2/M_1 =5$. The
    horizontal lines are the current cosmological bound of $Y_B$.}
        \begin{center}
        \includegraphics[width=16cm]{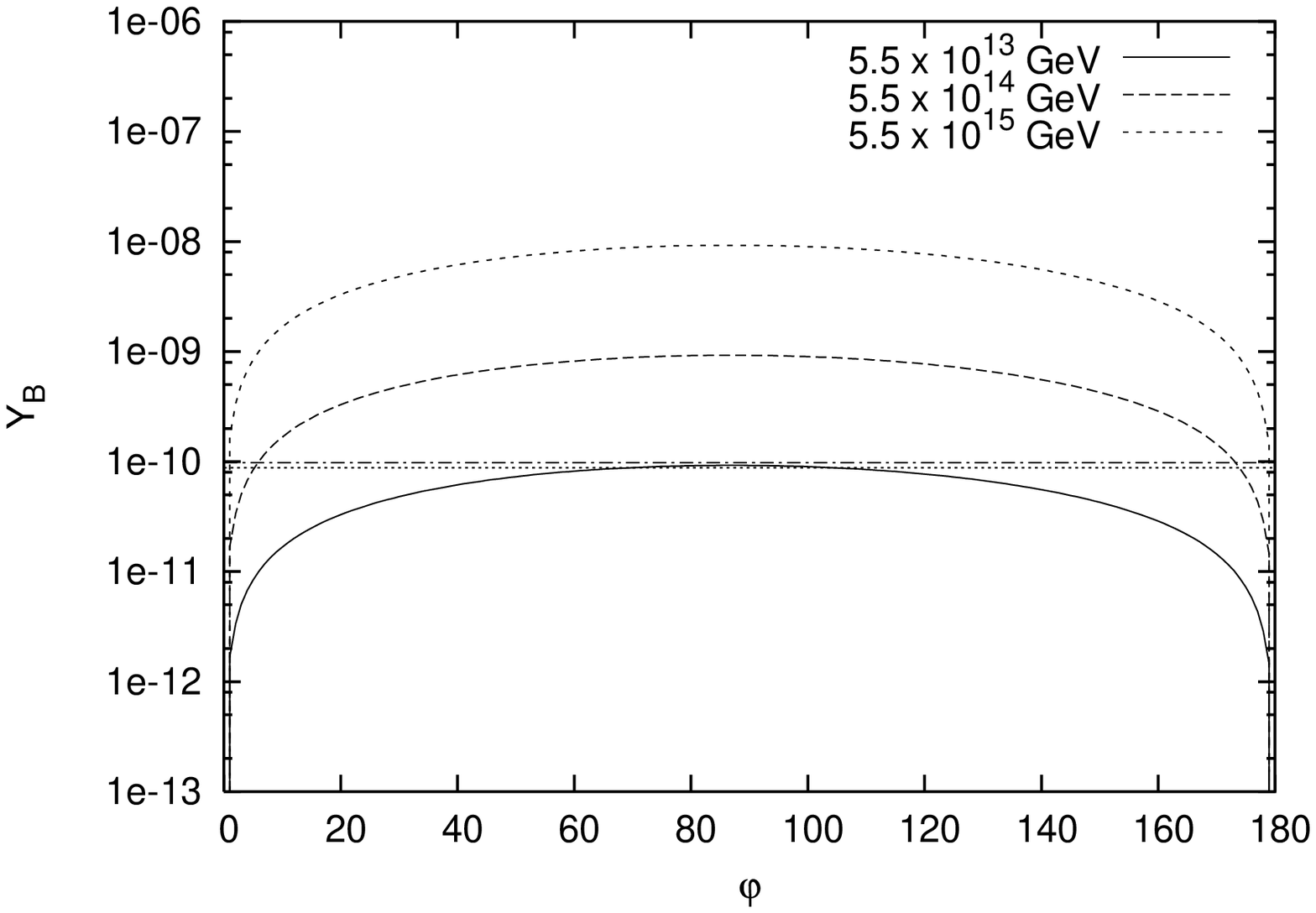}
        \end{center}
    \caption{$Y_B$ as a function of Majorana CP phase for case $IH$
    1-a, with various values of $M_1$ where $M_2/M_1 =5$}
\end{figure}

\begin{figure}
        \begin{center}
        \includegraphics[width=16cm]{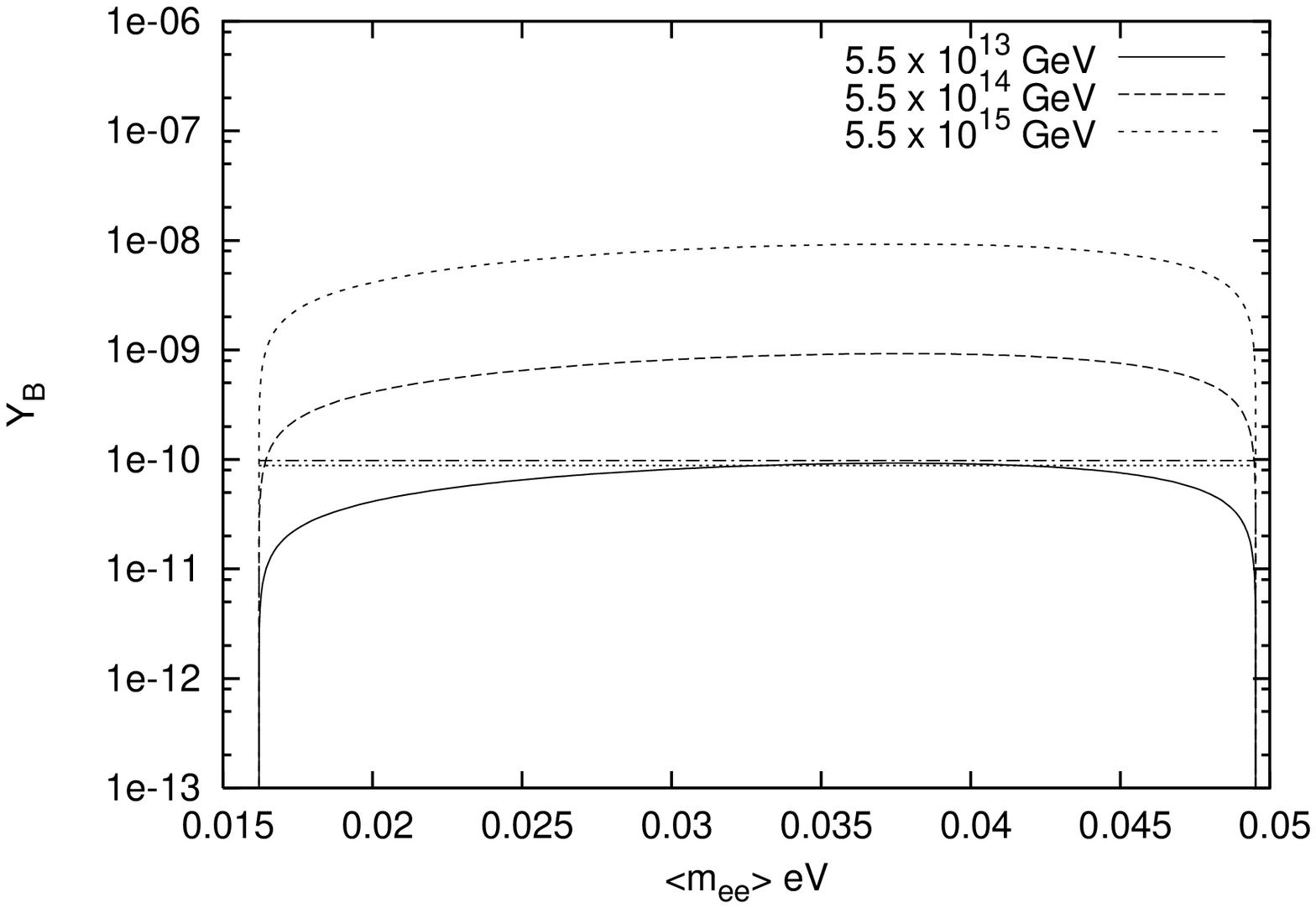}
        \end{center}
    \caption{$Y_B$ as a function of $|\langle m_{ee}\rangle |$ for
    case $IH$ 1-a, with various values of $M_1$ where $M_2/M_1 =5$}
        \begin{center}
        \includegraphics[width=16cm]{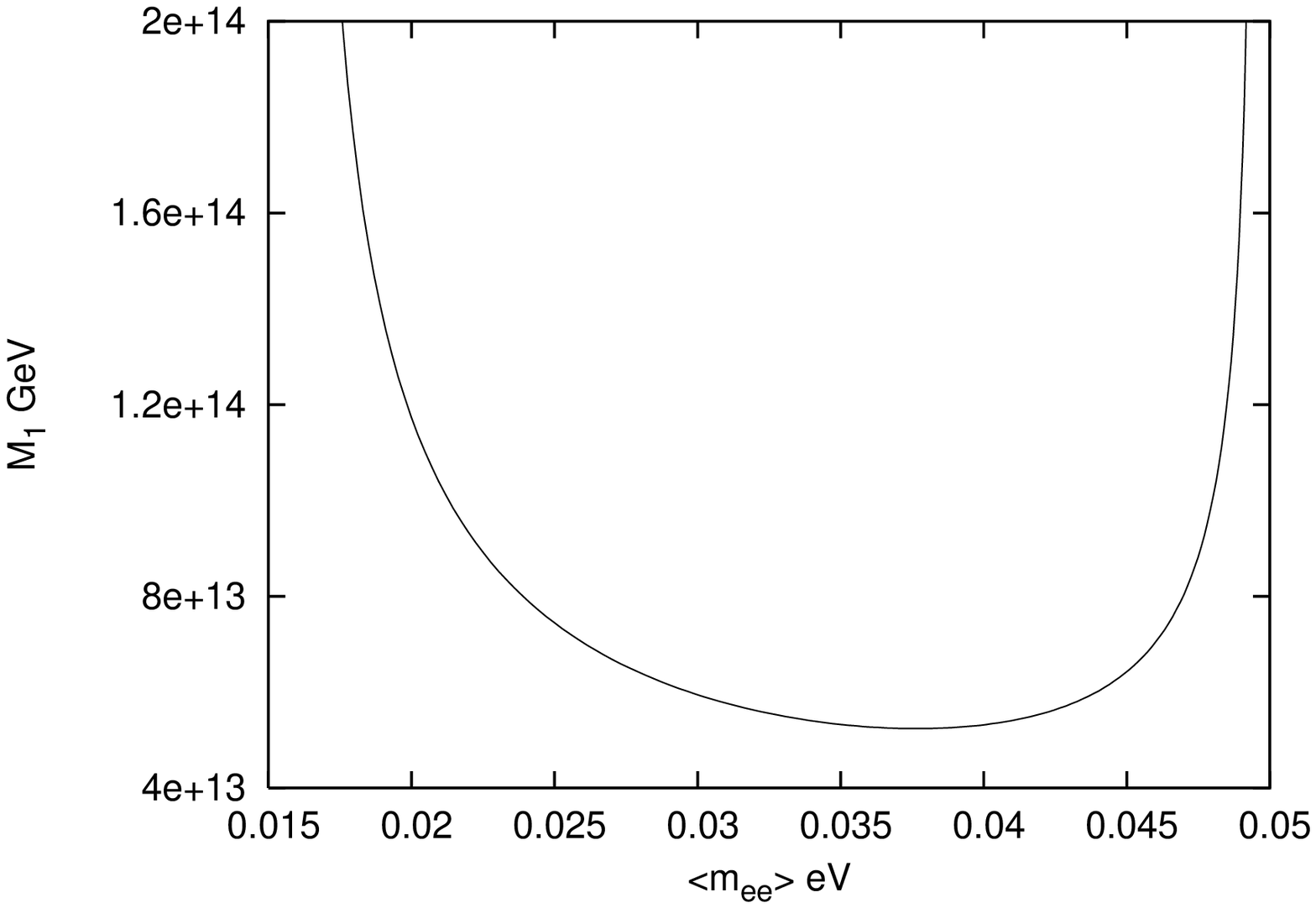}
        \end{center}
    \caption{The lower bound of $M_1$ as a function of $|\langle m_{ee}\rangle |$ for
    case $IH$ 1-a}
\end{figure}

\end{document}